\documentclass[11pt,a4paper,twocolumn]{article}

\usepackage{titlesec}
\usepackage{color}

\usepackage[utf8]{inputenc}
\usepackage[english]{babel}
\usepackage[T1]{fontenc} 

\usepackage{graphicx}
\graphicspath{{Images/}} 
\usepackage{eso-pic} 
\usepackage{subfig} 
\usepackage{caption} 
\usepackage{transparent}

\usepackage{amsmath}
\usepackage{amsthm}
\usepackage{bm}
\usepackage[overload]{empheq}  

\usepackage{tabularx}
\usepackage{longtable} 
\usepackage{colortbl}

\usepackage{algorithm}
\usepackage{algorithmic}

\usepackage[colorlinks=true,linkcolor=black,anchorcolor=black,citecolor=black,filecolor=black,menucolor=black,runcolor=black,urlcolor=black]{hyperref} 
\usepackage{cleveref}
\usepackage[square, numbers, sort&compress]{natbib} 
\usepackage{appendix}

\usepackage{enumitem}

\usepackage{amsthm,thmtools,xcolor} 
\usepackage{comment} 
\usepackage{fancyhdr} 
\usepackage{lipsum} 
\usepackage{tcolorbox} 
\usepackage{stfloats} 

\usepackage{amsfonts}

\newcommand{\bea}{\begin{eqnarray}} 
\newcommand{\eea}{\end{eqnarray}}

\usepackage{geometry}
\definecolor{bluePoli}{cmyk}{0.4,0.1,0,0.4}

\declaretheoremstyle[
  headfont=\color{bluePoli}\normalfont\bfseries,
  bodyfont=\color{black}\normalfont\itshape,
]{colored}

\theoremstyle{colored}

\newcounter{algsubstate}

\newcolumntype{L}[1]{>{\raggedright\let\newline\\\arraybackslash\hspace{0pt}}m{#1}}
\newcolumntype{C}[1]{>{\centering\let\newline\\\arraybackslash\hspace{0pt}}m{#1}}
\newcolumntype{R}[1]{>{\raggedleft\let\newline\\\arraybackslash\hspace{0pt}}m{#1}}

\setlist[itemize,1]{label=$\bullet$}
\setlist[itemize,2]{label=$\circ$}
\setlist[itemize,3]{label=$-$}
\setlist{nosep}

\newcommand\BackgroundPic{
	\put(230,358){
		\parbox[b][\paperheight]{\paperwidth}{%
			\vfill
			\centering
			\transparent{0.4}
			\includegraphics[width=0.5\paperwidth]{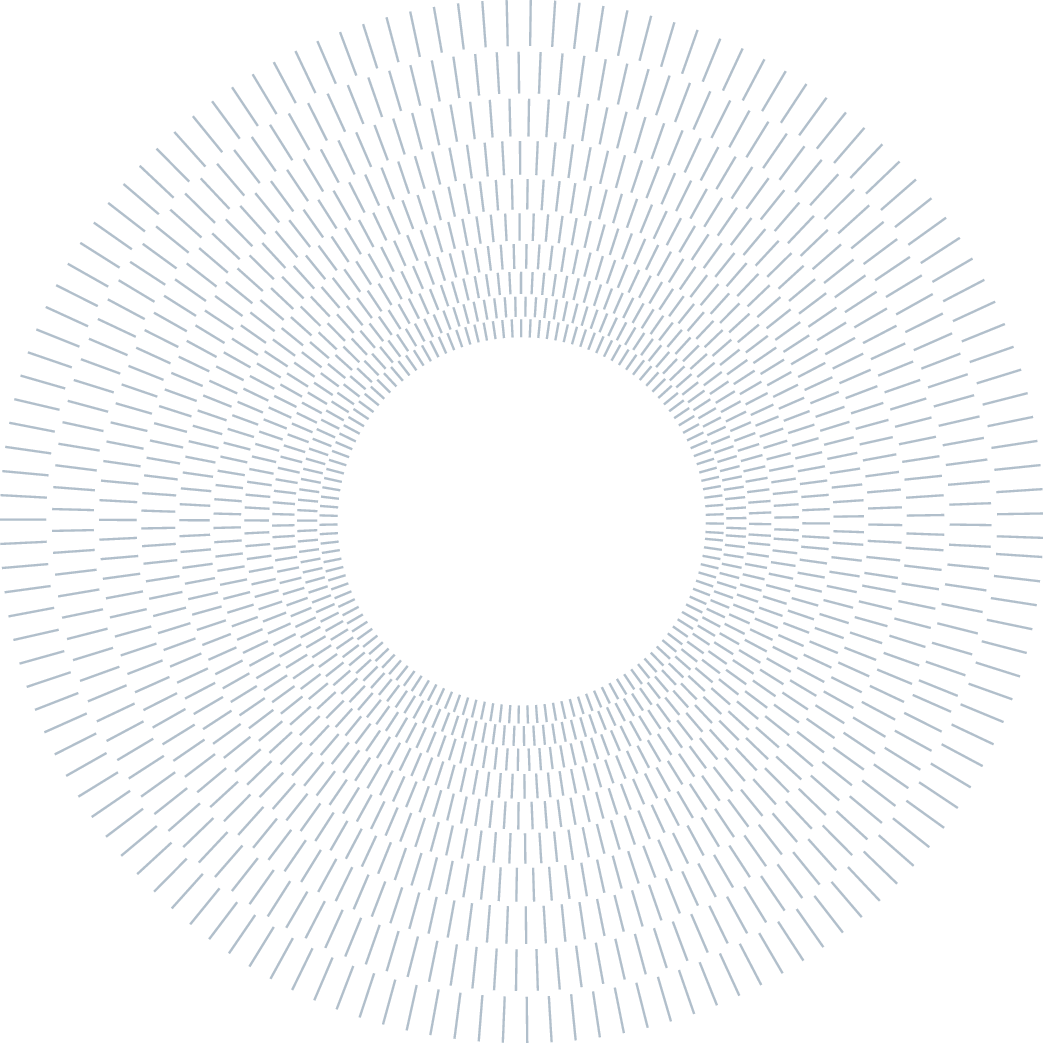}%
			\vfill
}}}

\titleformat{\section}
{\color{bluePoli}\normalfont\Large\bfseries}
{\color{bluePoli}\thesection.}{1em}{}
\titlespacing*{\section}
{0pt}{2ex}{1ex}

\titleformat{\subsection}
{\color{bluePoli}\normalfont\large\bfseries}
{\color{bluePoli}\thesubsection.}{1em}{}
\titlespacing*{\subsection}
{0pt}{2ex}{1ex}

\makeatletter
\patchcmd{\headrule}{\hrule}{\color{black}\hrule}{}{} 
\patchcmd{\footrule}{\hrule}{\color{black}\hrule}{}{} 
\makeatother

\renewcommand{\title}{VirtualRelativity: An Interactive Simulation of the Special Theory of Relativity in Virtual Reality}
\renewcommand{\author}{Alberto Boffi}
\newcommand{\course}{Computer Science and Engineering - Ingegneria Informatica}
\newcommand{\advisor}{Prof. Ezio Puppin}
\newcommand{\firstcoadvisor}{Maurizio Contran} 
\newcommand{\YEAR}{2023-2024}

\begin{document}


\twocolumn[{\begin{@twocolumnfalse}

\AddToShipoutPicture*{\BackgroundPic}

\hspace{-0.6cm}\includegraphics[width=0.6\textwidth]{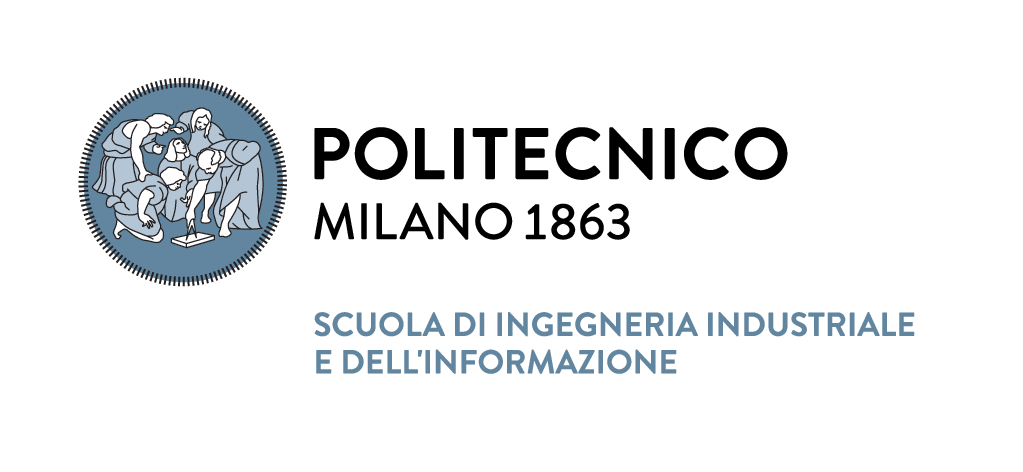}

\vspace{-1mm}
\fontsize{0.3cm}{0.5cm}\selectfont \bfseries \textsc{\color{bluePoli} Executive Summary of the Thesis}\\

\vspace{-0.2cm}
\Large{\textbf{\color{bluePoli}{\title}}}\\

\vspace{-0.2cm}
\fontsize{0.3cm}{0.5cm}\selectfont \bfseries \textsc{\color{bluePoli} Laurea Magistrale in \course}\\

\vspace{-0.2cm}
\fontsize{0.3cm}{0.5cm} \selectfont \bfseries Author: \textsc{\textbf{\author}}\\

\vspace{-0.4cm}
\fontsize{0.3cm}{0.5cm}\selectfont \bfseries Advisor: \textsc{\textbf{\advisor}}\\

\vspace{-0.4cm}
\fontsize{0.3cm}{0.5cm}\selectfont \bfseries Co-advisor: \textsc{\textbf{\firstcoadvisor}}\\

\vspace{-0.4cm}
\fontsize{0.3cm}{0.5cm}\selectfont \bfseries Academic year: \textsc{\textbf{\YEAR}}

\small \normalfont

\vspace{11pt}

\centerline{\rule{1.0\textwidth}{0.4pt}}

\vspace{15pt}
\end{@twocolumnfalse}}]

\thispagestyle{plain} 


\section{Introduction}
\label{sec:introduction}

The advent of \textbf{Virtual Reality} (\textbf{VR}) has heralded a new era in educational technology, offering immersive experiences that transform traditional learning paradigms. Among the most promising applications of VR is the visualization of complex scientific concepts \cite{burdea2003virtual} \cite{anthes2016state}. The \textbf{Special Theory of Relativity}, unveiled by \textbf{Albert Einstein} in the early 20th century, revolutionized our understanding of space, time, and motion \cite{einstein1905electrodynamics}. However, its counterintuitive consequences remain among the most challenging concepts for students and enthusiasts of physics to comprehend. Traditional learning methods, reliant on abstract mathematics and static visualizations, often struggle to fully engage the learner and translate theoretical concepts into concrete intuition \cite{jho2014literature} \cite{tanel2014student} \cite{selccuk2010addressing} \cite{scherr2002challenge}.
In this work, we introduce a novel solution in the form of a Unity package designed specifically to simulate the effects of special relativity. The package aims at delivering an exhaustive and accurate representation of the relativistic laws, featuring a carefully crafted interface that allows users to interact directly with these simulations. We also delve into the development of a VR application built on top of the package, that transports users into real-life scenarios where they can experiment with different relativistic effects.

\section{Background}

\subsection{Special Relativity}

In 1905, starting from Maxwell–Heaviside equations, \textbf{Albert Einstein} (14 March 1879 – 18 April 1955) formulated his concept of relativity in the \textbf{Special Theory of Relativity}. He based his theory on the postulates that we now know as \textit{postulates of special relativity}, asserting the absoluteness of the speed of light across all inertial reference frames \cite{einstein1905electrodynamics} \cite{einstein1912relativitat}. This consequence is incompatible with the Galilean relativity, leading to the Lorentz transformation. Given a stationary intertial reference frame \(S\) and a moving intertial reference frame \(S'\) travelling in one direction with speed \(v\), the Lorentz transformation describe how space and time of the two frames are related to each other by the relative motion (\ref{eq:lorentz_transformation_moving}) (\ref{eq:lorentz_transformation_static})  \cite{einstein1905electrodynamics} \cite{rao1988rotation} \cite{forshaw2014dynamics}.

\noindent\begin{minipage}{.5\linewidth}
\begin{equation}
    \label{eq:lorentz_transformation_moving}
    \begin{cases}
    x' & =\cfrac{x-vt}{\sqrt{1-\cfrac{v^{2}}{c^{2}}}} \\
    y' & =y \\
    z' & =z \\
    t'& =\cfrac{t-\cfrac{v}{c^{2}}\:x}{\sqrt{1-\cfrac{v^{2}}{c^{2}}}}
    \end{cases}
\end{equation}
\end{minipage}%
\begin{minipage}{.5\linewidth}
\begin{equation}
\label{eq:lorentz_transformation_static}
    \begin{cases}
    x & =\cfrac{x'+vt'}{\sqrt{1-\cfrac{v^{2}}{c^{2}}}} \\
    y & =y' \\
    z & =z' \\
    t & =\cfrac{t'+\cfrac{v}{c^{2}}\:x'}{\sqrt{1-\cfrac{v^{2}}{c^{2}}}}
    \end{cases}
\end{equation}
\end{minipage}

\vspace{\belowdisplayskip}

\subsection{Software and Hardware}

The choice for the software supporting the development fell into \textbf{Unity}, a cross-platform game engine launched in 2005 by \textbf{Unity Technologies}. It's now one of the leading environment for developing 2D, 3D, VR, and AR games for a variety of platforms, including PC, Mac, Linux, mobile devices and consoles \cite{haas2014history}. The user interaction is managed by the \textbf{XR Interaction Toolkit} (\textbf{XRI}), a high-level Unity package providing a framework for building cross-platform VR and AR experiences \cite{InteractionToolkit}. For the development and initial testing phases of this project, we employed a \textbf{Meta Quest 2}, a standalone VR headset developed by \textbf{Meta Platforms, Inc.}.

\section{System Overview}
\label{sec:system_overview}

\begin{figure}[H]
    \centering
    \includegraphics[width=0.25\textwidth]{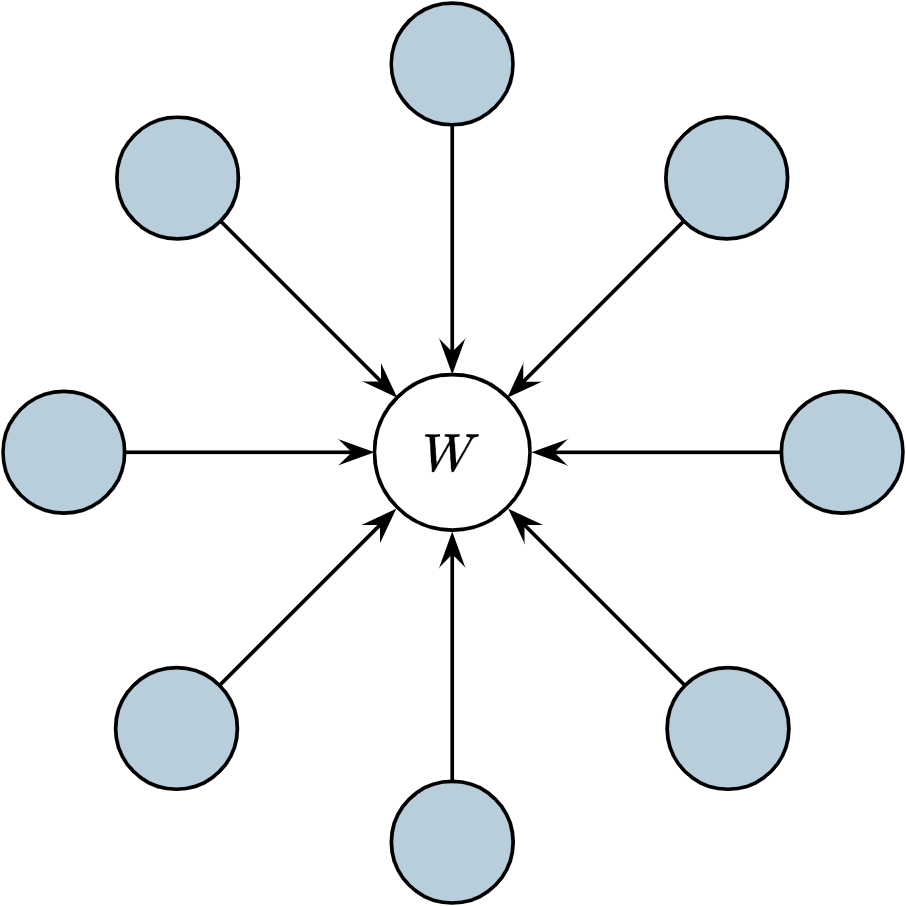}
    \caption{Graph representation of the optimized information flow.}
    \label{fig:optimized_inf_flow_vrm}
\end{figure}

We refer to the outcome of any application built using the package as the \textbf{virtual environment}. The virtual environment represents a model of our universe and it must be able to extract all the required information to faithfully represent the laws of special relativity. To ensure this, we structured the package over some axioms. The virtual environment can only contain three-dimensional objects. The kinematics of this objects (or, for convention, particles) is constrained in a one-directional motions in the x-direction. Moreover, due to the principle of inertial reference frames in the special theory of relativity, every particle moves with uniform speed with respect to any other particle. As each of them ascribes it own conception of length and flow of time to any other particle, conceptually we should store each speed as a set of relative values. However, this design requires quadratic storage space with respect to the number of particles, and linear time to account for the addition of a new particle within the environment. For this reason, we built the system upon the axiom asserting that each particle resides on a larger object that must adhere to the same behavioral constraints as all other objects within the environment. We call the respective inertial reference frame \textbf{World} (\(W\)), and we express all the speed relative to \(W\), reducing the computational complexities to \(S(n)=\Theta(n)\) and \(T(n)=\Theta(1)\) (Figure \ref{fig:optimized_inf_flow_vrm}).

At this point, we are left to analyze how a particle \(a\) can calculate the speed of another particle \(b\). For this purpose, we exploited the velocity composition law of special relativity, carefully translated to our problem. Thus, given the speeds \(v_{a}\) and \(v_{b}\) relative to \(W\), we can directly compute the speed \(v'_{b}\) of \(b\) relative to \(a\) (\ref{eq:vel_comp}) \cite{ungar1988thomas}.

\begin{equation}
    v\:'_{b}=\dfrac{v_{b}-v_{a}}{1-\cfrac{v_{a}v_{b}}{c^{2}}}
    \label{eq:vel_comp}
\end{equation}

As the relativistic effects are only visible at speeds close to the speed of light, when emulating everyday life situations we need to tamper with regular motions. To this purpose, we chosed to decrease the speed of light \(c\). Indeed, increasing the regular speeds of the particles would be an unnatural alternative, as well as difficult to both visualize and manage. This entail additional attention to the control over the speed of particles. In particular, as a consequence of special relativity, we must ensure that each particle moves in a time-like worldline (i.e. slower than the speed of light) \cite{fayngold2008special}.

\section{Package Architecture}

The architecture of the package is built upon three main components arranged in a two-level architecture. The \verb|MeshHandler| component operates at a low-level by leveraging the Unity's API to manange the meshes on the scene and the internal timing system, in an agnostic way with respect to the physical laws. On the top layer, the \verb|World| component represents the main engine of the package. Its roles encompasses the setup of the environment by means of the axioms governing the virtual environment, the orchestration of the motion of bodies and the application of the relativistic laws. The motion of bodies is managed in completely transparency to the user by converting all the provided speeds into the respective value relative to the observer, using the technique described in Section \ref{sec:system_overview}. Lastly, in the top-layer we also implemented the \verb|UIHandler| component, responsible for providing an ad-hoc interface aimed at each specific relativistic effect. The basic way the viewer can interact with the scene, for each portrayed phenomena, is by tampering with the speed of light through a \verb|UI.Sldier|.

\section{Technical Validation Setup}

\begin{figure}[H]
    \centering
    \includegraphics[width=0.45\textwidth]{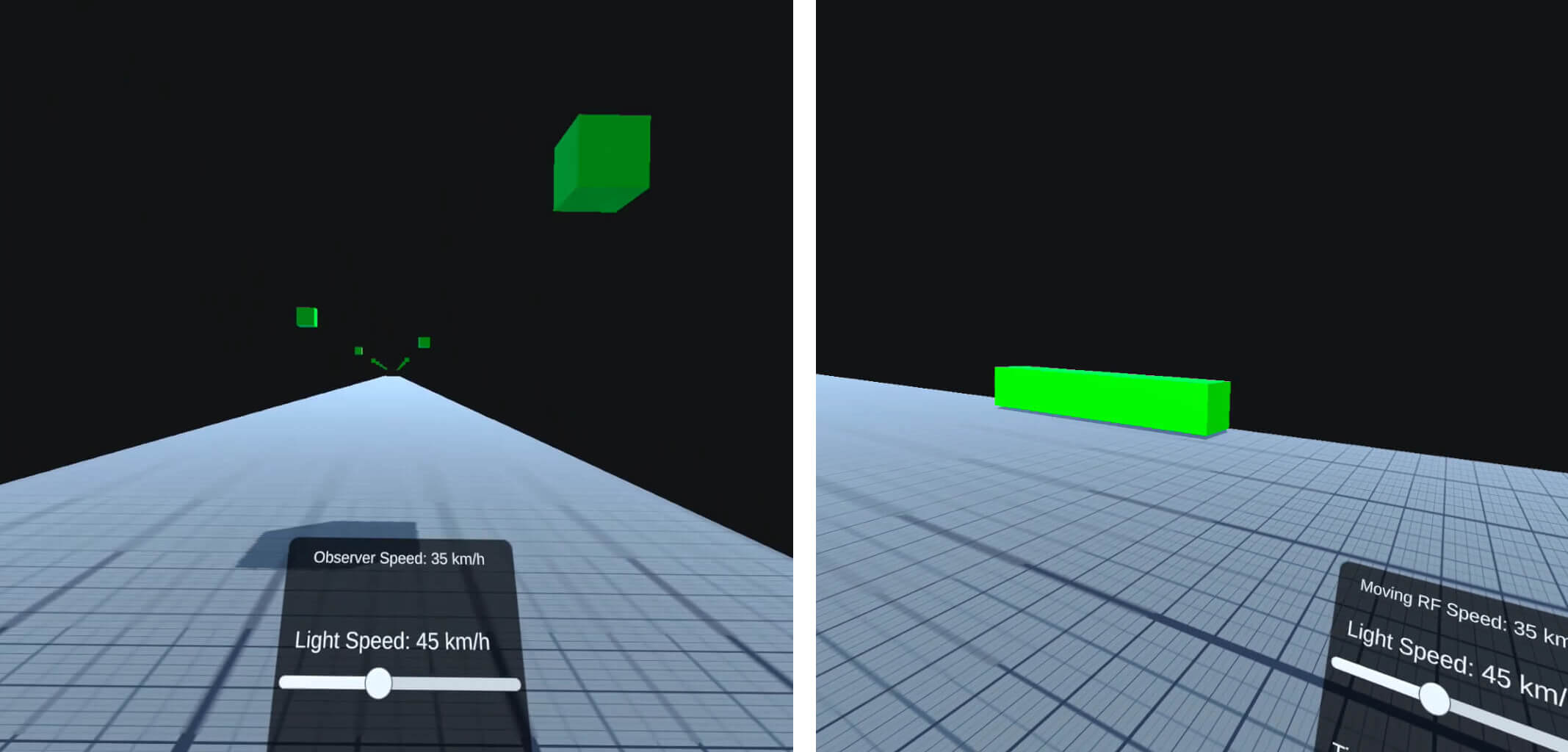}
    \caption{Examples of probe scenes.}
    \label{fig:probe_scenes}
\end{figure}

To ensure the robustness and accuracy of the package, we crafted focused test cases. The emphasis is in the implementation of \textbf{probe scenes}, each containing only the essential code and assets to represent a specific phenomenon (Figure \ref{fig:probe_scenes}). Probe scenes serve as isolated environments to rigorously evaluate the correctness and effectiveness of the package's underlying structure. From a techinical perspective, this approach allows for technical verification, code structure evaluation and performance optimization. The probe scenes also played a crucial role in refining how relativistic effects are presented to maximize intuitive comprehension. In this sense, we were able to tune the scenario, experimenting with the different kind of motions to better portrait the effect under study, while also testing the UI's ability to provide users with an effective interaction.

\section{Space Contraction}

\subsection{Overview}

We consider two reference frames \(S\) and \(S'\) in relative motion between each other. From the Lorentz transformations (\ref{eq:lorentz_transformation_moving}) (\ref{eq:lorentz_transformation_static}), both parties perceives a contraction in the spatial dimension of ther other reference frame that depends on the relative speed \(v\) (\ref{eq:length_contraction_stat}) (\ref{eq:length_contraction_mov}) \cite{einstein1905electrodynamics} \cite{shadowitz1988special}.

\noindent\begin{minipage}{.5\linewidth}
\begin{equation}
    \label{eq:length_contraction_stat}
    L'=L\sqrt{1-\cfrac{v^{2}}{c^{2}}} 
\end{equation}
\end{minipage}%
\begin{minipage}{.5\linewidth}
\begin{equation}
    \label{eq:length_contraction_mov}
    L=L'\sqrt{1-\cfrac{v^{2}}{c^{2}}}
\end{equation}
\end{minipage}

\vspace{\belowdisplayskip}

\subsection{Implementation}

The \verb|World| component applies the contraction effect by calculating the scaling factor and delegating the manipulation of the mesh to the \verb|MeshHandler|. The \verb|MeshHandler| calculates the low-level scaling factor needed to provide the desired relativistic effect given the local direction \(\vec{\sigma}\) corresponding to the world x-direction, the last scaling factor \(s_{p}\), the current scaling factor \(s\) and the current scale \(\vec{s}_{c}\) of the mesh (\ref{eq:space_contraction_implementation}).

\begin{align}
\begin{gathered}
    \label{eq:space_contraction_implementation}
    \vec{s}=(s-1)\cdot \vec{\sigma} + (1,1,1)^{\top}\\
    \vec{s}_{p}=(s_{p}-1)\cdot \vec{\sigma} + (1,1,1)^{\top} \\
    \vec{s}_{0}=\vec{s}_{c} \oslash \vec{s}_{p} \\
    \vec{s}_{new}=\vec{s}_{0} \odot \vec{s}
\end{gathered}
\end{align}

To design the interface, we needed to find a function \(\zeta: [0,1]\rightarrow \mathbb{F}\) to translate the position \(p\) of the slider's handle into the value of \(c\). A fundamental design goal is to ensure that equivalent handle displacements consistently produce equivalent magnitudes of change in relativistic effect. The lower bound of domain is set to  \(\zeta(0)=c_{min}=v_{max}{}^{+}\), where \(v_{max}\) is the maximum speed  of objects relative to the obsever. For the upper bound, instead of using the actual speed of light, a lower value is chosen to enhance the observable relativistic effects, addressing the issue of "relativistic saturation", where the high value of the speed of light and the sensitivity of the control slider limit the viewer's experience of the effect. To this purpose, we calculated the value of \(c_{max}\) by parametrizing it with the percentage \(k\) of space contraction manifestation (\ref{eq:cmax_slider_speed_light}). Fixed the extremes of the domain, the solution involves the user setting an initial speed of light (\(c_{0}\)) at an arbitrary point (\(p_{0}\)) on the slider, with a linear function \(\zeta_{0}\) applied to the left of \(p_{0}\) for direct control (\ref{eq:linear_slider_speed_light}), and an exponential function \(\zeta_{1}\) to the right to delay the increase of speed (\ref{eq:exponential_slider_speed_light}). Moreover, we added a "bullet time" effect (with no physical meaning) to allow viewers to analyze the effect more comfortably. The control over the speed of the moving reference frame is not needed since the magnitude of the effect is governed by a single degree of freedom expressed by the ratio \(\frac{v}{c}\) (\ref{eq:length_contraction_stat}) (\ref{eq:length_contraction_mov}). Moreover, this control doesn't provide beenfits in terms of user experience. 

\begin{equation}
    \label{eq:cmax_slider_speed_light}
    c_{max} = \sqrt{\frac{v_{max}^{2}}{2k-k^{2}}}
\end{equation}

\begin{equation}
    \label{eq:linear_slider_speed_light}
    c = \frac{c_{0}-c_{min}}{p_{0}} \cdot p + c_{min}
\end{equation}

\begin{equation}
    \label{eq:exponential_slider_speed_light}
    c = \frac{c_{max}}{\beta} \cdot \beta ^ {\;p},\:\:\beta=\left(\frac{c_{max}}{c_{0}}\right)^{\left(\frac{1}{1-p_{0}}\right)}
\end{equation}

For the test phase, we employed two symmetric kind of motions: a static observer framework (observer at rest relative to \(W\)) and a dynamic observer framework (observer in motion relative to \(W\)). The first one provided satisfactory results. The latter first required an analytical inspection: whenever there are multiple objects in motion relative to \(W\), we notice that the speed of the objects relative to the observer, derived as described in Section \ref{sec:system_overview} by means of the Lorentz transformation, accounts for both space contraction and time dilation (plus relativity of simultaneity). This means that, to isolate the space contraction effect, the only moving reference frame must be the one of the observer. Nevertheless, even in this situation, the motion scenario produced confusing results as intrinsically linked to paradoxes and issues.

\bigbreak

\section{Time Dilation}

\subsection{Overview}

We still consider two reference frames \(S\) and \(S'\) in motion with speed \(v\) relative to each other. In this case, we analyze the consequence of the Lorentz transformations (\ref{eq:lorentz_transformation_moving}) (\ref{eq:lorentz_transformation_static}) on the temporal dimension. In particular, both parties perceive a dilation of the time of the other reference frame (\ref{eq:time_dilation_stat}) (\ref{eq:time_dilation_mov}) \cite{adams2017relativity}.

\noindent\begin{minipage}{.5\linewidth}
\begin{equation}
    \label{eq:time_dilation_stat}
    \Delta t=\cfrac{\Delta t'}{\sqrt{1-\cfrac{v^{2}}{c^{2}}}} 
\end{equation}
\end{minipage}%
\begin{minipage}{.5\linewidth}
\begin{equation}
    \label{eq:time_dilation_mov}
    \Delta t'=\cfrac{\Delta t}{\sqrt{1-\cfrac{v^{2}}{c^{2}}}}
\end{equation}
\end{minipage}

\vspace{\belowdisplayskip}

\subsection{Implementation}

Implementing relativistic time dilation in a virtual environment differs from spatial contraction due to the less obvious nature of the effect. While spatial contraction is directly noticeable, time dilation requires observing events within the moving reference frame to perceive slowed temporal progression. The proposed system allow diverse events within the moving frame, focusing solely on controlling the rate of time flow within that frame. This is achieved by having each object of the environment to directly query the \verb|World| component for its specific \(\Delta t\) elapsed between frames. This design also addresses Unity's limitations in selectively manipulating time for individual objects.

About the interface, in this case we cannot allow the viewer to slow down time, as it would interfere with the true nature of the effect itself. Indeed, it's not strictly necessary to add additional commands beyond the control over the speed of light, provided that scenic setups allow the viewers to analyze the consequences of their input. Being the scaling factor for time the same of the one we've found for space, the interface design achieves the same results. We also decide to reconsider the control over the speed of the moving reference frame, but the handling on the conflicts with the parametrization of \(c\) causes confusing results on the viewer.

For the testing phase, we still considered two symmetric motions of the observer relative to \(W\). The static observer framework has been discarded due to the inherent issues in the implementation of nested-events and in the showcase of these events on moving objects. These issues are instead eliminated in the dynamic observer framework. In this case, we can avoid the effect of the space contraction in the composition of speed by simply refraining from deforming the mesh, isolating in this case the dilation of time. The results of the testing were even more satisfying than expected: the viewer was able to fully experience the time dilation phenomenon, comprehending and studying its effect based on the value of the speed of light.

\section{Relativistic Doppler Effect}

\subsection{Overview}

If we consider a wavelength \(\lambda'\) emitted by a source of light moving with speed \(v\) relative to an observer, the time dilation causes the received wavelength \(\lambda\) to be shifted with respect to \(\lambda'\). This shift is different in the two cases of convergent (\ref{eq:doppler_effect_towards}) and divergent (\ref{eq:doppler_effect_away}) motion between the source and the observer \cite{gill1965doppler}.

\noindent\begin{minipage}{.5\linewidth}
\begin{equation}
    \label{eq:doppler_effect_towards}
    \lambda = \lambda'\: \sqrt{\cfrac{c-v}{c+v}} 
\end{equation}
\end{minipage}%
\begin{minipage}{.5\linewidth}
\begin{equation}
    \label{eq:doppler_effect_away}
    \lambda = \lambda'\: \sqrt{\cfrac{c+v}{c-v}}
\end{equation}
\end{minipage}

\vspace{\belowdisplayskip}

In the convergent motion, the received wavelength convey toward the blue light of the visible spectrum. This phenomenon is known as \textbf{blueshift} \cite{kuhn2004quest}. Conversely, in the divergent motion light is deflected towards red, causing a \textbf{redshift} effect \cite{gray2008review}. As both the source and the observer are within inertial reference frames, the equations tell us that during both the blueshift and redshift phenomena the received wavelength remains constant \cite{gill1965doppler}.

\subsection{Light Representation In Computer Graphics}

There exists algorithms able to convert a given wavelength into the RGB encoding that simulates the corresponding perception of the human eye (a.k.a. spectral color) \cite{kuntzleman2016teaching}. However, we can't define a mixture of multiple wavelengths in terms of a single RGB color due to different obstacles, as the metamerism phenomenon \cite{foster2006frequency}, the role played by the light intensity \cite{kong2023dependence} and the nonlinearity of the color interpretation carried out by the human mind \cite{meessen1967simple}. Moreover, analyzing the opposite mapping, the gamut limitation of the RGB standard and the metamerism prevent us from converting a digital color into either a single or multiple wavelengths. This results impose a significant limitations that constrain us to portray the Doppler shift only on materials characterized by spectral color \cite{foster2006frequency} \cite{pascale2003review}.

\subsection{Implementation}

The implementation sees the \verb|MeshHandler| component returning the the mutual position between the observer and the source of light, which is combined with the speed \(v\) of the light source to determine the kind of relative motion. The \verb|World| component applies then the correct frequency shift. In turn, the \verb|MeshHandler| displays the shift by changing the color of the source's material accordingly.

For the interface, fixed the emitted wavelength and the speed of the moving reference frame, we still have an exponential relationship between the speed of light \(c\) and the received wavelength. For this reason, even in this case, the \verb|UIHandler| component must control the value of \(c\) by means of both a linear function \(\varphi_{0}:[0,p_{0}]\rightarrow \mathbb{F}\) and an exponential function \(\varphi_{1}:[p_{0},1]\rightarrow \mathbb{F}\). To define the values of \(\varphi_{0}(0)\) and \(\varphi_{1}(1)\), we want to provide a complete span of the received wavelength in the limits of the visible spectrum, represented by the interval \([\underline{\lambda}, \overline{\lambda}]\). For the upper limit of the domain, as \(\lim_{c\to \infty}\lambda = \lambda'\), we can set \(c_{max}=c_{real}\). For the lower bound, when the source moves towards the observer, the minimum value of \(c\) must led to \(\lambda=\underline{\lambda}\) (\ref{eq:c_min_de_towards}). Instead, when the source moves away from the observer, we need to impose \(\lambda=\overline{\lambda}\) (\ref{eq:c_min_de_away}). We derive the corresponding bounds, that must be merged into a single fixed value (\ref{eq:c_min_de}).

\begin{equation}
    \label{eq:c_min_de_towards}
    c^{t}_{min} = v\:\cfrac{\biggl[1+\biggl(\cfrac{\underline{\lambda}}{\lambda'}\biggr)^{2}\biggr]}{\biggl[1-\biggl(\cfrac{\underline{\lambda}}{\lambda'}\biggr)^{2}\biggr]} 
\end{equation}

\begin{equation}
    \label{eq:c_min_de_away}
    c^{a}_{min} = v\:\cfrac{\biggl[\biggl(\cfrac{\overline{\lambda}}{\lambda'}\biggr)^{2}+1\biggr]}{\biggl[\biggl(\cfrac{\overline{\lambda}}{\lambda'}\biggr)^{2}-1\biggr]}
\end{equation}

\begin{equation}
    \label{eq:c_min_de}
    c_{min}=\max\{c^{t}_{min}, c^{a}_{min}\}
\end{equation}

This results can be used to construct the functions \(\varphi_{0}\) and \(\varphi_{1}\) as per the standard.

The test case resembles the scenario set up for the space contraction effect. The outcome was excellent: the viewer was able to fully analyze the phenomena and parametrize it in comfortable settings.

\section{Validation Outcomes and Performance Evaluation}

The validation process has successfully confirmed the package potential to meet the key requirements of achieving a faithful simulation of special relativity and providing a suitable  UI that is both accessible and efficient. We also proved the ability to enable developers to create VR experiences without requiring extensive coding expertise in complex physics calculations. This is achieved by the architecture of library offered by the package, which abstracts the complex mathematics into a user-friendly API. Moreover, our probe scene-based evaluation yielded valuable insights into the performance of the package, revealing its efficiency across both CPU and memory usage. Different profiling metrics revealed several positive indicators. We first noticed that the processes consuming the bulk of computational resources (i.e. \verb|Gfx.WaitforpresentOnGfxThread| or \verb|Camera.Render|) were system scripts independent of the developed package \cite{Technologies_2023}, enabling us to boast efficient resource utilization. Memory profiling provided further positive results. The package demonstrates a modest memory footprint of approximately 570 MB, with the graphics drivers and the profiler itself accounting for a substantial portion of overall memory allocation.

\section{VirtualRelativity: A Practical Demonstration}

\begin{figure}[H]
    \centering
    \includegraphics[width=0.4\textwidth]{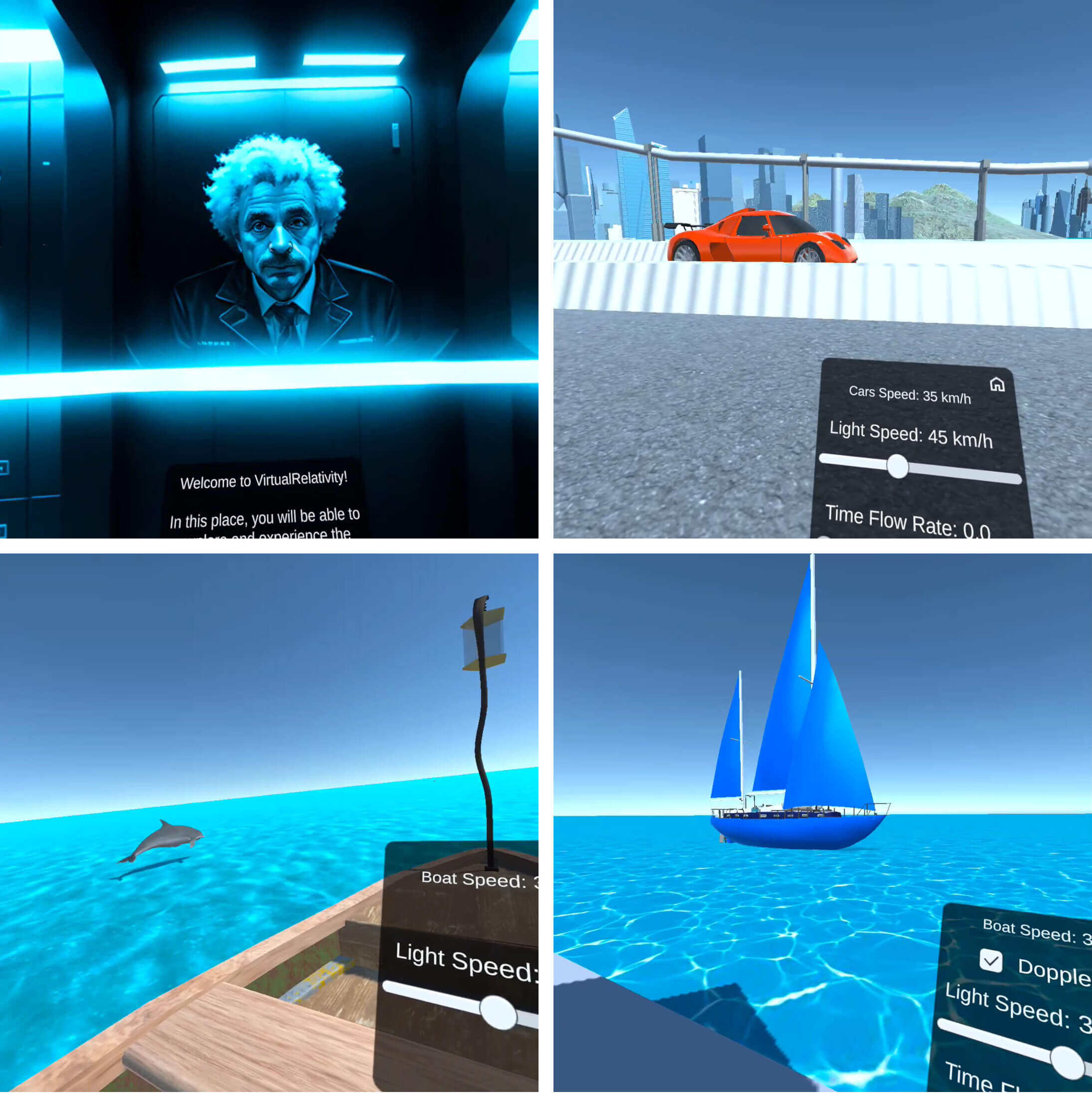}
    \caption{Screenshots from VirtualRelativity.}
    \label{fig:application_screens}
\end{figure}

In the pursuit of featuring our work, we developed a concrete application, \textbf{VirtualRelativity}, that underscores the package's technical versatility and its potential to transform the understanding of special relativity within virtual spaces (Figure \ref{fig:application_screens}). At its core, VirtualRelativity features a main menu housed within the interior of a stylized spaceship, where users can choose to be teleported to several real-life scenario, each designed to illustrate a specific relativistic effect. The user is guided inside the application by a background voice. The application employes asynchronous scene loading to mitigate performance overhead in complex environments.
The first experience places users on the Brooklyn Bridge in New York City, where they can experience the space contraction effect on moving vehicles. This vehicles are spaced at a fixed temporal interval, ensuring that only one of them is visible within the user's viewport at any given time. Moreover, a continuous experience is guaranteed by cyclically repositioning vehicles at the start of their journey.
The time dilation scene transports users to an open ocean aboard a moving boat, providing a dynamic environment to observe the relativistic effect on leaping dolphins. Based on the previously analyzed approach, the package simulates the boat's forward motion by subtly manipulating the surrounding ocean environment. The illusion of an infinite sea is achieved through a coordinated repositioning of multiple Plane GameObjects representing segments of the ocean. The dolphins exhibit a natural parabolic motion pattern. The horizontal (x-axis) movement is directly managed by the package's built-in motion management funcionality, which seamlessly incorporates the appropriate time dilation factor. The vertical (y-axis)  motion is governed by the standard rules of projectile motion, where the time value is provided by the package, ensuring the impact of relativistic time dilation is accurately reflected in the dolphins' trajectories.
The Doppler effect scene leverages the same fundamental motion settings established in the space contraction scene, where users finds themselves within a lakeside house, observing boats passing in perpetual motion.
Initially, the camera is centered in the environment by moving the \verb|XR Origin| component depending on the position and orientation of the \verb|MainCamera|. For the rest of the experience, the player can move using the controllers or by physical movements, while the application moves the UI panel to mantain the interface's accessibility.

\section{Empirical Experimentation Insights}

Following the development of VirtualRelativity, we employed a multifaced experimentation phase, including expert evaluation and cross-disciplinary feedbacks. Moreover, the application will be utilized by students enrolled in the inaugural "Relativity" course for master's degree students at the Politecnico di Milano. Offered in the 2023/2024 academic year, this course is taught by Professor Ezio Puppin with a specific emphasis on special relativity. To provide the experience, we will leverage the Politecnico's existing laboratory, equipped with 14 Meta Quest 2 headsets, that provides an infrastructure enabling streamlined distribution and wired execution of the application in the form of a PC executable.

\section{Conclusions}

In this work, we presented the design, implementation and evaluation of a virtual reality software package tailored to the rigorous simulation of special relativity. The results demonstrate the successful creation of a package that accurately models relativistic laws, prioritizes efficiency and performance, and incorporates a thoughtfully crafted user interface to enhance understanding and interactivity for the user. Verification of these objectives has been carried out via a rigorous testing phase and through the development of a specific application, VirtualRelativity. This application has led to valuable opportunities for disseminating an understanding of special relativity among students, initially at the Politecnico di Milano, with potential expansion to other educational institutions in the future.

\bibliography{bibliography.bib} 

\end{document}